# FUSE observations of molecular hydrogen on the line of sight towards HD 141569A.

C. Martin-Zaïdi[1], M. Deleuil[1], T. Simon[2], J.-C. Bouret[1], A. Roberge[3], P.D. Feldman[4], A. Lecavelier des Etangs[5], and A. Vidal-Madjar[5]

[1] Laboratoire d'Astrophysique de Marseille, CNRS-Université de Provence, BP 8, 13376 Marseille Cedex 12, France
   e-mail: `Claire.Martin@oamp.fr`
[2] Institute for Astronomy, University of Hawaii, 2680 Woodlawn Drive, Honolulu, HI 96822, USA
[3] Carnegie Institution of Washington, Washington, DC, USA
[4] Department of Physics and Astronomy, JHU, Baltimore, MD, USA
[5] Institut d'Astrophysique de Paris, France

**Abstract.** We present an analysis of the Far Ultraviolet Spectroscopic Explorer (*FUSE*) spectrum of HD 141569A, a transitional object known to possess a circumstellar disk. We observe two components of gas at widely different temperatures along the line of sight. We detect cold $H_2$, which is thermalized up to $J = 2$ at a kinetic temperature of $\sim 51$ K. Such low temperatures are typical of the diffuse interstellar medium. Since the line of sight to HD 141569A does not pass through its disk, it appears that we are observing the cold $H_2$ in a low extinction envelope associated with the high Galactic latitude dark cloud complex L134N, which is in the same direction and at nearly the same distance as HD 141569A. The column densities of the higher J-levels of $H_2$ suggest the presence of warm gas along the line of sight. The excitation conditions do not seem to be consistent with what is generally observed in diffuse interstellar clouds. The observed radial velocity of the gas implies that the UV spectral lines we observe are likely interstellar in origin rather than circumstellar, although our absorption line study does not definitely rule out the possibility that the warm gas is close to the star. The discovery of such warm gas along the line of sight may provide evidence for turbulent phenomena in the dark cloud L134N.

## 1. Introduction

HD 141569A is a high Galactic latitude B9.5V star located at 108±6 pc from the Sun (Merín et al. 2004). It is a transitional object passing from the pre-main sequence Herbig star stage into a zero-age main sequence (ZAMS) star. Its age has been estimated at more than 10 Myr from *Hipparcos* data by van den Ancker et al. (1998) and at 5±3 Myr by Weinberger et al. (2000), while a recent study by Merín et al. (2004) fixed it at 4.71±0.3 Myr. HD 141569A is not associated with any dense cloud or reflexion nebula, and is the highest mass (2.0 $M_\odot$; Merín et al. 2004) member of a triple system of stars (Weinberger et al. 2000), its two companions being M2 and M4 type stars. Moreover, HD 141569A is known to possess an extended ($R = 400$ AU) circumstellar disk (Weinberger et al. 1999; Augereau et al. 1999; Mouillet et al. 2001; Clampin et al. 2003), which has a region of depleted material or a gap, as a double ring structure (Weinberger et al. 1999; Dent et al. 2005). The inclination angle to the line of sight estimated for the disk is 51°±3° (Weinberger et al. 1999). The spectral energy distribution (SED) of HD 141569A presents a strong excess at far-IR wavelengths but a low excess at near-IR wavelengths, $\lambda \leq 5\mu$m (Malfait et al. 1998), which implies that most of the dust in the central part of the disk ($\leq 50$ AU) near the star has been dissipated. In addition, Brittain & Rettig (2002) have demonstrated from their CO data that the inner disk surrounding the star is past the early phase of accretion and planetesimal formation, and that most of the gas and dust has been cleared out to a distance of more than 17 AU. However, Merín et al. (2004) have shown that the SED of HD 141569A is easily reproduced with a model of a slightly flared disk at a distance of 0.24 AU from the star, which is consistent with a substantial amount of residual gas in the disk.

A high Galactic latitude dark cloud complex, referred to as L134N, lies close to the star (Sahu et al. 1998). The extinction towards the centre of L134N is at least $A_v \sim 6$ (Laureijs et al. 1991) and may exceed $A_v = 10$. The cloud is surrounded by an extended low extinction ($A_v < 1$) envelope (Juvela et al. 2002). It is associated with the CO core MBM37 (Magnani et al. 1985). Its distance is estimated to be 110±10 pc (Franco 1989), which is very similar to that of the star. HD 141569A clearly does

*Send offprint requests to*: C. Martin-Zaïdi

not lie within the dark cloud itself. Despite controversies, Sahu et al. (1998) concluded from their ultra-high resolution observations at optical wavelengths that the star is behind the complex. They demonstrated that two different media are observed in absorption towards HD 141569A: the first one may correspond to an outer region of the dark cloud complex L134N at a radial velocity about +20.1 km s$^{-1}$ in the star's rest frame, and the second one more likely corresponds to a region close to the star, and has a radial velocity of +5.1 km s$^{-1}$ in the star's rest frame.

In this paper, we present our analysis of the *FUSE* spectrum of HD 141569A covering the spectral range from 905 to 1187 Å ($R \simeq 15\,000$). Section 3 deals with our results about the gaseous media we observed along the line of sight, especially our analysis of the H$_2$ lines. An extensive discussion about the location of the detected gas is given in Sect. 4, and our conclusions are presented in Sect. 5.

## 2. Observation and data reduction

HD 141569A was observed on April 3, 2004, with the *FUSE* 30″×30″ LWRS aperture at a resolution about 15 000. The data were processed with the 3.0.7 version of the *FUSE* pipeline CalFUSE. The total exposure time is 7283 seconds, split into 4 subexposures. The signal-to-noise ratio (S/N) per pixel is about 3 to 5 in the co-added spectra in each detector channel. These spectra were rebinned in wavelength by a factor 5 in order to increase the S/N per pixel to 15 at 1150 Å, without degrading the resolution.

HD 141569A is separated from its two companions HD 141569B and HD 141569C, M2V and M4V-type respectively, by less than 9″ in projection (Weinberger et al. 2000). These companions could fall in the *FUSE* aperture, but the extremely low FUV fluxes of M-type stars most likely do not affect our observations of HD 141569A.

In the HD 141569A *FUSE* spectrum, the stellar continuum is well detected down to wavelengths shortward of 1040 Å. Numerous circumstellar and/or interstellar lines of molecular hydrogen and atomic species are seen in absorption in the stellar spectrum. The analysis of those lines has been performed using the Owens profile fitting procedure (Hébrard et al. 2002; Lemoine et al. 2002). Owens treats all the lines of one or more chemical species simultaneously, throughout several spectral windows. By spectral windows we mean a series of small subspectra centered on absorption lines to be analyzed, whose widths are a few angstroms. The stellar continuum is locally fitted with polynomials. We generally set the degrees of these polynomials in a range from 1 to 4 (see Fig. 1 and Fig. 2), but it could be as high as 14 if necessary. Owens creates a synthetic spectrum for a set of physical parameters: radial velocity, turbulent and thermal rms velocity, column densities. It calculates on each point of a grid in wavelength space the total optical depth by summing the optical depths associated with each line of absorbing species. Each optical depth is the result of the calculation of a Voigt function with the proper physical parameters, and the synthetic spectrum is further convolved with the instrumental line spread function. Fitting is done via $\chi^2$ minimization until finding the best fit model. Column densities were derived from fitting of unsaturated and/or damped lines. The radial velocities were measured from the unsaturated lines and the line widths (*b*) from the saturated lines. Examples of Owens's use are illustrated in Fig. 1 and Fig. 2.

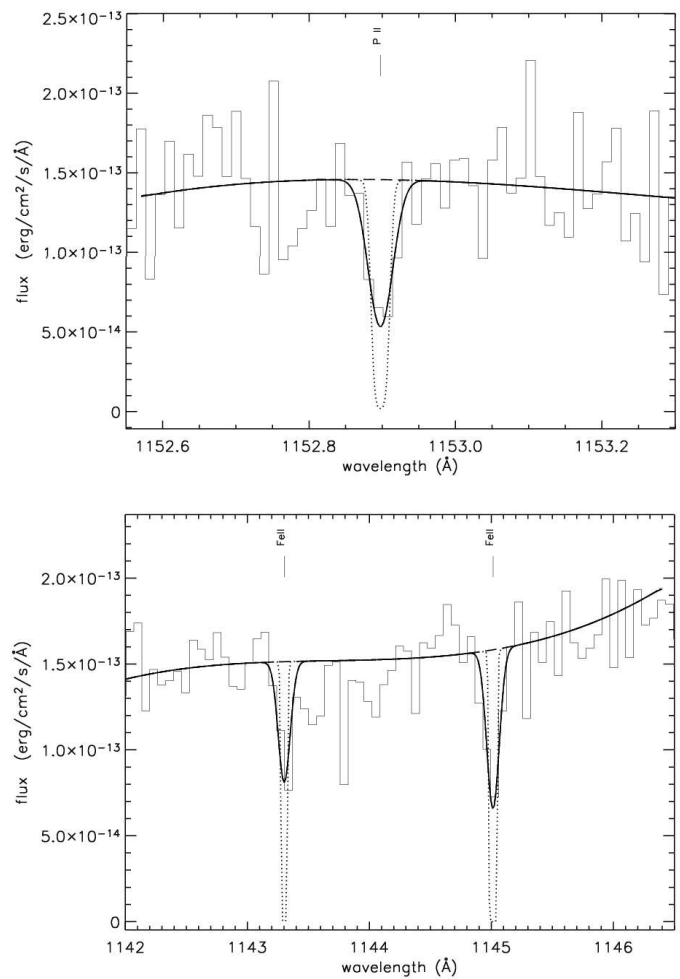

**Fig. 1.** Examples of fits of atomic lines with the Owens fitting procedure. The observed spectrum is not corrected for reddening. Stellar continuum: dashed line; Intrinsic line profile: dotted line; Resulting profile convolved with the line spread function: thick line. At the top of the plot, fit of the resonance absorption line of P II near 1152.8Å, and at the bottom of the plot, fit of two resonance absorption lines of Fe II near 1143 and 1144 Å respectively.

No hot emission lines or wind lines are observed in the spectrum. This lack of activity in the FUV domain is consistent with the evolutionary stage of the star between the pre-main sequence phase and the ZAMS. In some of the subexposures of the SiC 2b segment, very weak emission lines of the O VI $\lambda\lambda$1032-1038 resonance doublet are apparent but are due to the Sun (for an example of Solar contamination of the SiC channel, see Lecavelier des Etangs et al. 2004).

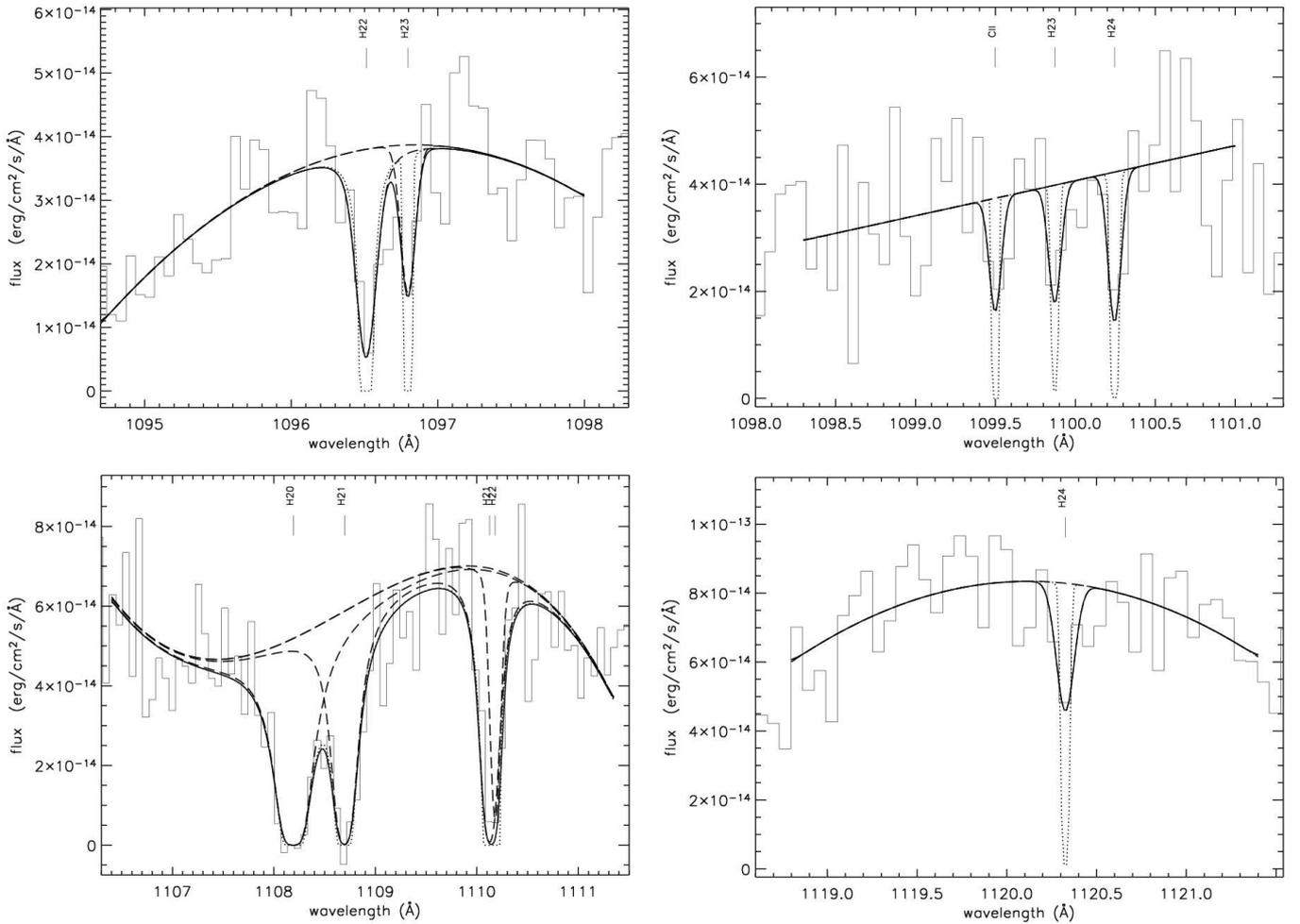

**Fig. 2.** Examples of fits of molecular hydrogen absorption lines with the OWENS fitting procedure. The H$_2$ levels are marked H20 for ($v = 0, J = 0$), H21 for ($v = 0, J = 1$) etc... Note that atomic species can be fitted simultaneously with H$_2$ lines. In some cases, it is not necessary to fit atomic lines with the H$_2$ ones, and some of the features in the plots are not identified but are present in other *FUSE* spectra (see text, Sect. 3.2.1).

## 3. Results

### 3.1. Atomic species

In the HD 141569A *FUSE* spectrum, numerous absorption lines of Ar I, N I, N II, Cl I, P II and Fe II corresponding to electronic transitions from the ground level as well as fine structure levels have been identified. We used FUV wavelengths and oscillator strengths tabulated by Morton (2000) to analyse those lines with OWENS (see Fig. 1). All these lines are observed at nearly the same radial velocity. The results we obtained from our fitting procedure, with 2-$\sigma$ error bars, are given in Table 1.

### 3.2. Molecular gas

#### 3.2.1. Molecular hydrogen

We identified H$_2$ electronic lines corresponding to the rotational levels $J = 0$ to $J = 4$ in the ground vibrational state ($v = 0$). All the $J = 0$ to $J = 3$ lines are detected at a confidence level from 3$\sigma$ to 7$\sigma$ while the weaker $J = 4$ lines are detected at a 2$\sigma$ to 3$\sigma$ confidence level. Higher rotational J-levels and vibrational levels are beyond the detection limit, because of the low S/N of the data (less than 10 in the corresponding spectral ranges). The detection sigma has been estimated by computing the uncertainty in the flux in the co-added spectrum, wavelength by wavelength (pixel by pixel).

We have fitted all the H$_2$ lines simultaneously in several spectral windows. When necessary, we also add the nearby atomic lines. The transitions of H$_2$ have been modelled using data tabulated in Abgrall et al. (1993a,b), Abgrall et al. (2000) and Balakrishnan et al. (1999). Examples of the simultaneous fit of several H$_2$ lines are shown in Fig 2. The column densities we measured for each energy level of H$_2$ are given in Table 2. All the H$_2$ lines corresponding to the lower J-levels ($J = 0 - 2$) are damped or saturated. The radial velocity was measured from the unsaturated lines and is $-25^{+2.41}_{-1.83}$ km s$^{-1}$ in the observed rest frame, which is nominally heliocentric. We assume that the lower J-levels have the same velocity as the higher J-levels. This velocity is similar to the value we measured for all the atomic species. From the saturated lines, we determined $b = 5.3^{+0.51}_{-0.36}$ km s$^{-1}$.

Fig. 3 presents the excitation diagram corresponding to the H$_2$ column densities towards HD 141569A derived from our *FUSE* spectrum. This plot shows that two temperatures are needed to fit the measurements. The H$_2$ is thermalized up to $J = 2$ with a kinetic temperature of 51±6 K. These excitation conditions are typical of those generally observed in the diffuse interstellar medium (Gry et al. 2002). Such a low temperature gives evidence for the presence of cold H$_2$ along the line of sight towards HD 141569A.

**Table 1.** Atomic lines detected in the *FUSE* spectrum of HD 141569A. The velocities are in the pipeline's rest frame (observed spectrum) and are heliocentric. For the saturated lines, we measured the velocity of the centroid; for those lines the error bars on the velocity are symetric (not calculated by a $\chi^2$ minimization). The radial velocities and line widths have been determined using respectively all the unsaturated and saturated lines of a given ion. "sat" and "unsat" are respectively for saturated and unsaturated lines. $-^a$ line not fitted because of the very low S/N ratio of the data; $-^b$ Doppler parameter $b$ not measured because of the unsaturated state of the line.

| Element | Wavelength Å | Energy level (cm$^{-1}$) | Line State | Radial Velocity (km s$^{-1}$) | N (cm$^{-2}$) | $b$ (km s$^{-1}$) |
|---|---|---|---|---|---|---|
| Ar I | 1048.2198 | 0.00 | $-^a$ | $-27.0\pm3$ | $>7.30\times10^{14}$ | $5.6^{+0.42}_{-0.25}$ |
|  | 1066.6599 | 0.00 | sat |  |  |  |
| FeII | 1063.1764 | 0.00 | unsat | $-24.28^{+1.89}_{-2.75}$ | $9.77^{+4.56}_{-0.94}\times10^{15}$ | $2.34^{+0.25}_{-0.79}$ |
|  | 1096.8770 | 0.00 | sat |  |  |  |
|  | 1121.9749 | 0.00 | unsat |  |  |  |
|  | 1143.2260 | 0.00 | unsat |  |  |  |
|  | 1144.9379 | 0.00 | sat |  |  |  |
| FeII* | 1068.3456 | 384.79 | unsat | $-24.28^{+1.89}_{-2.75}$ | $5.21^{+3.65}_{-1.25}\times10^{13}$ | $2.34^{+0.25}_{-0.79}$ |
|  | 1096.6073 | 384.79 | unsat |  |  |  |
|  | 1147.4092 | 384.79 | unsat |  |  |  |
|  | 1148.2773 | 384.79 | unsat |  |  |  |
| FeII** | 1071.5842 | 667.68 | unsat | $-24.28^{+1.89}_{-2.75}$ | $4.25^{+4.36}_{-2.65}\times10^{13}$ | $2.34^{+0.25}_{-0.79}$ |
|  | 1151.1458 | 667.68 | unsat |  |  |  |
| FeII*** | 1148.0789 | 862.61 | unsat | $-24.28^{+1.89}_{-2.75}$ | $2.77^{+1.12}_{-0.89}\times10^{13}$ | $2.34^{+0.25}_{-0.79}$ |
|  | 1153.2719 | 862.61 | unsat |  |  |  |
| N I | 1134.1653 | 0.00 | sat | $-24.0\pm2$ | $>3.16\times10^{14}$ | $6.80^{+2.11}_{-3.05}$ |
|  | 1134.4149 | 0.00 | sat |  |  |  |
|  | 1134.9803 | 0.00 | sat |  |  |  |
| N II | 1083.9937 | 0.00 | $-^a$ | $-22.0\pm4$ |  | $-^a$ |
| N II* | 1084.5659 | 48.67 | $-^a$ | $-22.0\pm4$ |  | $-^a$ |
|  | 1084.5841 | 48.67 | $-^a$ |  |  |  |
| N II** | 1085.5511 | 130.80 | $-^a$ | $-22.0\pm4$ |  | $-^a$ |
|  | 1085.7096 | 130.80 | $-^a$ |  |  |  |
| P II | 1152.8180 | 0.00 | unsat | $-25.37^{+1.56}_{-1.25}$ | $1.00^{+0.65}_{-0.67}\times10^{13}$ | $-^b$ |
| P II* | 1149.9580 | 164.90 | unsat | $-25.37^{+1.56}_{-1.25}$ | $1.03^{+0.74}_{-0.56}\times10^{13}$ | $-^b$ |
| P II** | 1153.9951 | 469.12 | unsat | $-25.37^{+1.56}_{-1.25}$ | $1.24^{+0.70}_{-0.81}\times10^{13}$ | $-^b$ |
| Cl I | 1097.3691 | 0.00 | unsat | $-25.12^{+1.64}_{-1.13}$ | $3.12^{+0.82}_{-1.53}\times10^{14}$ | $-^b$ |
|  | 1099.5229 | 0.00 | unsat |  |  | $-^b$ |

The column densities of the $J = 3$ and $J = 4$ levels are much greater than expected for a kinetic temperature of 51 K. Therefore, in addition to the cold gas along the line of sight, there must also be a second component of much warmer H$_2$. However, the ratio of the column densities in the higher J-levels ($J = 3 - 4$) gives an unphysical fit in Fig. 3 since log(N$_{J=3}$/g$_3$)<log(N$_{J=4}$/g$_4$). This inconsistency probably due to the large uncertainty in the column density we find for the $J = 4$ level. Within the errors on the column densities for the $J = 3$ and $J = 4$, we can only state that the lower limit on the temperature of the warm gas is 270 K.

Since the $J = 4$ lines are very weak and fall in very noisy regions of the spectrum, we undertook a statistical analysis to estimate the robustness of our detection. Three $J = 4$ lines of $H_2$ are included in the fits in different spectral windows. All the $J = 4$ features correspond to the same radial velocity and column density. In addition, we have also fitted the stellar continuum without $H_2$ lines in the spectral windows of interest. The presence of $J = 4$ lines significantly improves the $\chi^2$ of the fit to the data in comparison with the fit without $H_2$. We concluded that our detection of the $J = 4$ lines is real at the 99% confidence level. However, in order to reduce the uncertainty on the temperature of the higher J-levels, we proceeded as follow: we assumed an excitation temperature of the high J-levels of 300K, which is comparable to the lower limit given by the excitation diagram, and which is an upper limit of what is generally observed in the diffuse interstellar medium. Using the column density of the $J = 3$ level, we derived the column density of the $J = 4$ level expected for such a temperature. We found that this column density should be $5.0 \times 10^{14}$ cm$^{-2}$; note that this value correspond to the lower error bar on our best fit value. Assuming this column density, we then computed the $\chi^2$ probability distribution taking into account the total number of degrees of freedom used in our fitting procedure. We found that such column density can be excluded at more than 98%. We used the same method with different excitation temperatures between 300 K and 1500 K. When computing the $\chi^2$ probability distribution, the level of confidence increases with the temperature. For example, a temperature of 1000 K is rejected only at 70%, compared to the 98% rejection of a temperature of 300 K. This suggests that the temperature of the high J-levels is much higher than 300 K, which is hotter than what is observed in the diffuse interstellar medium.

**Table 2.** Column densities of the different energy levels of $H_2$. $g$ is the statistical weight of each level. "cold" and "warm" $H_2$ are distinguished from the excitation diagram (see Fig. 3). "sat" and "unsat" are respectively for saturated and unsaturated lines. Only the fitted lines are listed.

| J (v=0) | g | Energy level (cm$^{-1}$) | Wavelength Å | Line state | N (cm$^{-2}$) |
|---|---|---|---|---|---|
| 0 | 1 | 0.00 | 1092.2020 | damped | $1.20^{+0.75}_{-0.23}\times10^{20}$ |
|   |   |      | 1108.1390 | damped |   |
| 1 | 9 | 118.49 | 1092.7390 | damped | $9.12^{+4.58}_{-6.09}\times10^{19}$ |
|   |   |       | 1094.0590 | damped |   |
|   |   |       | 1108.6440 | damped |   |
|   |   |       | 1110.0740 | damped |   |
| 2 | 5 | 354.39 | 1066.9070 | sat | $1.58^{+6.16}_{-0.93}\times10^{16}$ |
|   |   |        | 1094.2510 | damped |   |
|   |   |        | 1096.4460 | sat |   |
|   |   |        | 1110.1310 | damped |   |
|   |   |        | 1112.5081 | unsat |   |
| 3 | 21 | 705.50 | 1067.4850 | sat | $1.10^{+1.81}_{-0.79}\times10^{16}$ |
|   |   |        | 1096.7310 | sat |   |
|   |   |        | 1099.7950 | unsat |   |
|   |   |        | 1112.5940 | unsat |   |
|   |   |        | 1115.9070 | unsat |   |
| 4 | 9 | 1168.53 | 1100.1689 | unsat | $6.02^{+8.08}_{-5.15}\times10^{15}$ |
|   |   |         | 1116.0250 | unsat |   |
|   |   |         | 1120.2610 | unsat |   |
| cold $H_2$ | J=0-2 |  |  |  | $2.11^{+1.21}_{-0.83}\times10^{20}$ |
| warm $H_2$ | J=3-5 |  |  |  | $1.70^{+2.62}_{-0.94}\times10^{16}$ |

As shown in Fig 2, numerous absorption "features" are observed in the neighbourhood of the $J = 4$ lines and some of them correspond to atomic species. To confirm these features, we compared the HD 141569A spectrum with that of HD 100546 (Lecavelier des Etangs et al. 2003; Deleuil et al. 2004) and that of AB Aurigæ (Roberge et al. 2001; Martin et al. 2005). The HD 100546 *FUSE* spectrum (a particularly deep 22 ks exposure) serves as an excellent template for comparison as the S/N ratio is extremely high in the spectral windows containing the $J = 4$ lines. All the features detected in the HD 141569A spectrum match those observed in the other spectra. Moreover, in the HD 100546 and AB Aurigæ spectra some features near the $J = 4$ $H_2$ lines are not identified (not tabulated?). These features are also present in the HD 141569A spectrum and thus do not correspond to noise features.

We also note that two very weak lines corresponding to the $J = 5$ rotational level may also be present in the *FUSE* spectrum at a marginal detection level of just 1.5 $\sigma$. If the detection of these lines is real, then the column density given by Owens is about $2.69^{+11.8}_{-2.33}\times10^{15}$ cm$^{-2}$. The excitation temperature obtained from simultaneously fitting the column densities of the $J = 3-5$ levels is then 945±189 K. This result favours the presence of warm $H_2$ along the line of sight towards HD 141569A.

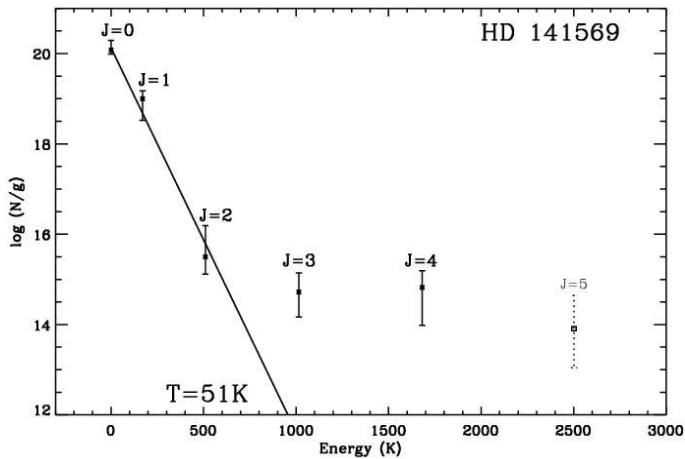

**Fig. 3.** Excitation diagram for $H_2$ in HD 141569A. The column densities of the lower J-levels are consistent with a kinetic temperature of 51 K while the higher J-levels give a lower limit on their excitation temperature about 270 K, but our statistical analysis shows that the temperature is likely much higher than that (see text, Sect. 3.2.1). This diagrams reveals both cold and warm components along the line of sight. Note that the $J = 5$ level is plotted with dotted lines because of the high uncertainty on its detection.

### 3.2.2. Carbon monoxide

The strongest CO bands that can be detected in the *FUSE* spectral range are $^{12}$CO bands: the E-X band at 1076 Å and the C-X band at 1087 Å. Other CO lines, including those of $^{13}$CO, also fall in this spectral domain but are much weaker.

No CO absorption band is observed in the *FUSE* spectrum of HD 141569A. This is consistent with the observations of Caillault et al. (1995) who showed that the star falls just outside the outermost CO contours that trace MBM37, the CO core in L134N. From the *FUSE* spectrum we set an upper limit on the total column density of CO of $\leq 2 \times 10^{14}$ cm$^{-2}$. Assuming this upper limit on the column density for CO, we find $N(CO)/N(H_2) \leq 10^{-6}$, which is consistent with the typical ratio in the diffuse interstellar medium (Ferlet et al. 2000; André et al. 2004).

## 4. Discussion

The radial velocity of the star in the heliocentric rest frame is -6.4 km s$^{-1}$ (Frisch 1987). In the observed *FUSE* spectrum, we measured a radial velocity about -25 km s$^{-1}$ for $H_2$ and the different atomic species. Nevertheless, the *FUSE* data suffer from an uncertainty in the wavelength calibration which can be as large as 30 km s$^{-1}$ in the heliocentric rest frame. It is thus necessary to adjust the absolute wavelength scale.

As a first step, we used a photospheric model with the appropriate stellar parameters (Merín et al. 2004), in order to establish the absolute wavelength calibration of our spectrum (for details see Bouret et al. 2003). Due to the presence of relatively broad interstellar and/or circumstellar (IS/CS) lines superimposed on the photospheric lines, this method gives a very high uncertainty (± 30 km s$^{-1}$) on the wavelength calibration. With such an uncertainty, the estimates on the radial velocities of the different species are not accurate enough to determine the location of the gas along the line of sight.

In a previous study, using optical spectroscopy data at ultra high resolution (∼940 000), Sahu et al. (1998) identified two absorption components along the line of sight: the first one, probably close to the star, has a radial velocity of +5.1 km s$^{-1}$ in the star's rest frame (-1.3 km s$^{-1}$ heliocentric, or -15 km s$^{-1}$ LSR) and a second one at +20.1 km s$^{-1}$ in the star's rest frame (+13.7 km s$^{-1}$ heliocentric, or 0 km s$^{-1}$ LSR), which most likely corresponds to an outer region of the interstellar dark cloud L134N. The first component is detected from the Ca II K line only, while the second one is observed in Ca II K, Na I D1 and D2, K I, and CH lines.

Since the formation of CH is predicted to be controlled by gas-phase reactions with $H_2$, the CH molecule should be a good tracer of $H_2$ (Federman 1982; Mattila 1986; Somerville & Smith 1989). We therefore assumed that our $H_2$ gas is located in the same region and is at the same velocity as the CH gas observed at optical wavelengths, that is, we set $v_{H_2} = v_{CH} = +20.1$ km s$^{-1}$ in the star's rest frame. With the resolution of *FUSE* the residual uncertainty on this velocity is about ±2 km s$^{-1}$. Note that the discrepancy (38.7 km s$^{-1}$) between the velocities of far ultraviolet lines and the radial velocities of the optical lines is higher than the estimated uncertainty in the absolute scale of the *FUSE* data (30 km s$^{-1}$). At the present time, we cannot explain this phenomenon.

As shown earlier, all the atomic species have a radial velocity very close to that of $H_2$. In that case, we do not observe any lines at a velocity of about +5 km s$^{-1}$ in the star's rest frame. On the other hand, by shifting our spectrum in order to fix the radial velocity of the detected gas at +5 km s$^{-1}$, no atomic lines are identified at any other velocity. This would mean that we do not detect any atomic lines from the +20 km s$^{-1}$ component and is thus inconsistent with the detection of several atomic lines in the optical domain. This suggests that we only detect in the FUV domain the component corresponding to an outer region of the dark cloud L134N at +20 km s$^{-1}$.

The total cold $H_2$ column density we derive from the *FUSE* spectrum, $N(H_2) = 2.11^{+1.21}_{-0.83} \times 10^{20}$ cm$^{-2}$, is in agreement with the $N(H_2)$ estimate of Sahu et al. (1998) for the component at +20 km s$^{-1}$. Indeed, they derived $N(H_2) = 1.5 \pm 0.9 \times 10^{20}$ cm$^{-2}$ from their CH observations of L134N. In addition, the low excitation temperature of the lower J-levels of $H_2$ ($J \leq 2$) along the line of sight towards HD 141569A is comparable to what is generally observed in interstellar molecular clouds (Gry et al. 2002; Rachford et al. 2002). Moreover, numerous lines of atomic species are observed at the same velocity as the $H_2$ and no lines of very ionized or excited species are present in the spectrum, which is consistent with the chemistry of the diffuse interstellar medium (ISM). Assuming that the populations of the first two levels of Fe II follow a Boltzmann law, for example, we found that their column densities correspond to a kinetic temperature of ∼ 76 K. Using the same method, we found a temperature of ∼ 92 K from the P II column densities. These temperatures are of the same order of magnitude as what we found for $H_2$ and imply that these elements are located in the same cold, dense gaseous medium.

From the radio observations at 21 cm obtained in the Leiden/Dwingeloo H I survey, Sahu et al. (1998) determined the column density of H I in L134N. They found $N(H I) = 5.03 \times 10^{20}$ cm$^{-2}$. Using this value, the fractional abundance given by

$$f(H_2) = \frac{2N(H_2)}{2N(H_2) + N(H\,I)}$$

is therefore $0.46^{+0.11}_{-0.12}$ for the region of the dark cloud L134N we observe in absorption. This value of $f(H_2)$ is typical of that observed for the diffuse component associated with high Galactic latitude molecular clouds (Penprase 1993). All these results favour an interstellar origin of the detected gas.

However, we cannot rule out the presence of a second component at +5 km s$^{-1}$ in the H$_2$ lines. The H$_2$ lines corresponding to energy levels up to $J = 3$ are broad and could be a blend of the two components detected optically at high resolution. In that case, the resolution of *FUSE* may be too low to separate these components. Our $\chi^2$ tests do not allow us to discriminate the presence of two different gaseous media in the broad damped lines of H$_2$. Nevertheless, if we assume these two components are present on the line of sight and adopt the column density of Ca II determined by Sahu et al. (1998), we should easily identify the component at +5 km s$^{-1}$ in our *FUSE* spectrum. Indeed, Ca II is less abundant in the diffuse ISM than N I, N II and P II, and thus the component at +5 km s$^{-1}$ should be observed in these atomic lines, which is not the case. The $J = 4$ lines of H$_2$ and atomic lines are clearly too narrow to be blends of two different velocity components.

Therefore, our analysis of the *FUSE* spectrum of HD 141569A gives no evidence for the presence of the circumstellar gaseous component at +5 km s$^{-1}$. In this context, an excitation temperature $\gg 300$ K for the higher J-levels ($J = 3-4$) of H$_2$ and the relatively large values of the Doppler parameter $b$ are quite surprising. These physical conditions of the warm gas are incompatible with the classical chemistry of the diffuse ISM. According to recent results, the heating of the gas and an excitation temperature around or higher than 1000 K could be explained by the dissipation of turbulence, in shocks or in vortices, within a layer of the dark cloud (Falgarone et al. 2005). It is possible that the warm gas we observe corresponds to the transition layer between the hot and cold regions at the edge of the cloud. In addition, if the lines were only thermally broadened, the $b$ value should scale with the atomic and molecular mass. Here, the $b$ value of H$_2$ is similar to those of the atomic species and thus this suggests a non-thermal (i.e. turbulent) broadening.

In summary, the excitation diagram of H$_2$ obtained from the *FUSE* data shows the presence of two media along the line of sight. Firstly, the lower J-levels ($J = 0 - 2$) most likely correspond to a cold inner region in the dark cloud L134N, and secondly, the excitation of the higher J-levels ($J = 3 - 4$) is possibly the result of turbulence in a thin external layer of the dark cloud. From the warm region of L134N, we only observe the $J = 3$ to $J = 4$ levels because of the very low column densities of the corresponding lower J-levels ($J = 0 - 2$). Indeed, the column densities of these levels should be at least 100 times lower than those we observed for the cold H$_2$. Being in the same cloud, these cold and warm components cannot be distinguished from their radial velocities, which are nearly similar. This implies that L134N probably has the sufficient physical conditions to explain the relatively high excitation temperature of the warm H$_2$ without invoking the presence of gas very close to the star.

## 5. Conclusion

In the *FUSE* spectrum of HD 141569A, we have identified numerous absorption lines of H$_2$ and atomic species at the same radial velocity (at ~20 km s$^{-1}$ velocity resolution), suggesting that all these species are located in the same place. The excitation of the lower J-levels of H$_2$ and the lack of highly ionized and excited atomic species show that we probably observe the lightly reddened diffuse outer region surrounding the L134N dark cloud complex (Juvela et al. 2002). These results confirm those obtained from high resolution observations towards HD 141569A at optical wavelengths concerning the L134N component (Sahu et al. 1998). In addition, our *FUSE* data suggest the presence of warm excited gas that had not been observed before. While the excitation conditions of the lower J-levels of H$_2$ are typical of conditions in the diffuse interstellar medium, those of the higher J-levels favour an interpretation in terms of a turbulent region in the dark cloud.

Moreover, our *FUSE* data show no evidence of CS gas close to the star. In the case of HD 141569A, the CS gas is supposed to be located in a disk. The observations by Brittain & Rettig (2002) and Brittain et al. (2003) of the CO emission lines from the disk suggested that all the gas and dust have been cleared out to a distance of more than 17 AU from the star. The inclination angle of the disk is about 51°, and our line of sight to the star does not pass through the disk. Contrary to what has been previously observed for HD 100546 and HD 163296 (Lecavelier des Etangs et al. 2003), two other Herbig stars with CS disks having inclination angles similar to that of HD 141569A, we do not observe for the latter a part of the surface of the inner disk. The nondetection of CS gas implies that there is no remnant of a CS envelope and all the gas has had time to collapse into a flat or very slightly flared disk. This is consistent wih the $^{12}$CO observations at 345.796 GHz by Dent et al. (2005). Those authors indeed modelled their data with an opening angle for the disk of 1.0°. Our results confirm the evolutionary status of HD 141569A emphasized by the CO observations.

Our FUV observations of HD 141569A raises the question of the physical conditions of formation and excitation of the gas. In this context, new optical spectroscopic obervations at high resolution would be necessary to observe other absorption features, especially lines of CH$^+$ (near 3957Å and 4232Å) if present. Indeed, the formation of the CH$^+$ molecule through the chemical reaction C$^+$ + H$_2$ needs a temperature about 4500 K to occur (Gredel 1997) and is generally associated with the presence of a warm component of H$_2$. Thus, the CH$^+$ molecule is a probe of excited media. The observation of CH$^+$, if present, provides the most direct and most definitive evidence to better constrain the excitation of the higher J-levels of H$_2$.

*Acknowledgements.* This research is based on observations made with the NASA-CNES-CSA Far Ultraviolet Spectroscopic Explorer. FUSE is operated for NASA by the Johns Hopkins University under NASA contract NAS5-32985. HD 141569A was included in the *"Circumstellar Disks Program"* Q319 of the *FUSE* guaranteed time (P.I. Deleuil M.). We warmly thank T. Rettig for his comments on the paper. We also thank E. Falgarone, C. Gry and J. Le Bourlot for the fruitful discussions and comments, and B. Godard at J.H.U for reprocessing data for us. TS acknowledges support of FUSE Guest Observer Grant NNG04GO40G from NASA and thanks the Laboratoire d'Astrophysique de Marseille (LAM) for the gracious hospitality that was extended to him during his visit there.